\documentclass[reprint,preprintnumbers,nofootinbib,amsmath,amssymb,aps]{revtex4-2}

\usepackage{dcolumn,bm}
\usepackage{xcolor}

\begin{document}

\preprint{YITP-24-90, IPMU24-0032}

\title{Violations of energy conservation in Horava-Lifshitz gravity: a new ingredient in the dark matter puzzle}

\author{Paolo M Bassani}
\thanks{paolo.bassani22@imperial.ac.uk}
\affiliation{%
 Theoretical Physics Group, The Blackett Laboratory, Imperial College, Prince Consort Rd., London, SW7 2BZ, United Kingdom}%

\author{João Magueijo}
\thanks{j.magueijo@imperial.ac.uk}
\affiliation{%
 Theoretical Physics Group, The Blackett Laboratory, Imperial College, Prince Consort Rd., London, SW7 2BZ, United Kingdom}%

\author{Shinji Mukohyama}
\thanks{shinji.mukohyama@yukawa.kyoto-u.ac.jp}
\affiliation{%
Center for Gravitational Physics and Quantum Information, Yukawa Institute for Theoretical Physics, Kyoto University, 606-8502, Kyoto, Japan}%
\affiliation{%
Kavli Institute for the Physics and Mathematics of the Universe (WPI), The University of Tokyo Institutes for Advanced Study, The University of Tokyo, Kashiwa, Chiba 277-8583, Japan}%
 
\date{}

\begin{abstract}
We investigate the interplay between Horava-Lifshitz (HL) gravity and more general theories where the local Hamiltonian constraint is lost, for example due to the time variability of the Lagrangian (e.g. via its parameters) where time is defined on a foliation according to a prescription mimicking Lambda and 4-volume time in unimodualr gravity. In one direction we subject the multitude of parameters in HL to this variability game, mimicking RG flow in a cosmological setting. In the opposite direction, we examine the evolution on the left-over Hamiltonian should the HL algebra of constraints be still applicable, rather than the algebra of General Relativity being restored. Within the projectable theory, the non-vanishing Hamiltonian can be reinterpreted as a pressureless fluid, resulting in essentially the same phenomenologies at macroscopic scales as in the standard cold dark matter paradigm. At high energies and short distances, however, unlike in theories with similar variability based on GR, violations of stress-energy tensor conservation persist, and these are computed here for the full class of projectable HL models. The phenomenological implications are examined: remarkably the driven solution resulting from these energy conservation violations is shown to be the attractor of the system during a free-fall collapse as far as the backreaction is negligible. When the backreaction is taken into account, the driven solution is expected to play an important role towards our understanding of microscopic caustic avoidance, which is one of the most significant issues in many alternatives to particle dark matter scenarios. 
\end{abstract}

\maketitle

%\tableofcontents

\section{Introduction}

Since its first proposal~\cite{HL}, Horava-Lifshitz (HL) gravity has attracted continued attention (see for example~\cite{shinji_review} for an early appraisal and~\cite{HL-recentreview} for a recent review). One of the theory's central features is degradation of 4D space-time symmetry to 3D spatial diffeomorphisms, plus a global time redefinition invariance. This loss of symmetry entails the loss of the local Hamiltonian constraint and the emergence of a new degree of freedom. It was realized as early as in~\cite{shinji} that, in the low energy regime of the theory, this could be interpreted as a matter fluid with properties similar to those of cold dark matter (CDM). Although it was never explicitly stated in the early literature, such dust matter should be seen as the low energy expression of the extra degree of freedom of the theory, resulting from shrinking the symmetry from 4D to 3D diffeomorphism invariance. The much feared scalar graviton actually becomes non-relativistic matter, potentially being part of the puzzle of the missing matter in our Universe. 

This issue was brought into focus with the recent analysis of a much larger class of theories with an identical symmetry breaking pattern~\cite{geoCDM,nongeoCDM}. In general the breaking could come about from interactions with a foliation (dubbed global or Machian interactions), for example if there were to be evolution in the
laws of physics~\cite{evol} in terms of global time variables generalizing 4-volume time in unimodular theories~\cite{Magueijo:2021rpi,Magueijo:2021pvq,unimod,unimod1,UnimodLee1,alan,daughton,sorkin1,sorkin2,Bombelli,UnimodLee2,pad,pad1,lomb,vikman,vikman1,vikaxion}. In such theories, even after the symmetry breaking interactions switch off and 4D diffeomorphism symmetry is restored, there is permanent damage. The usual algebra of constraints of General Relativity
would feel the symmetry breaking effects for ever after the end of the symmetry breaking phase in the form of a leftover violation of the Hamiltonian constraint. 

In this paper we conflate these two theoretical frameworks in two ways. First, in Section~\ref{Sec:HL} (after briefly reviewing projectable HL theory in Section~\ref{HL-review}) we examine how the formalism for defining time and time dependence can be applied to the multitude of parameters defining HL theory. In particular such formalism could be used to simulate the RG flow of these parameters in a cosmological setting. Second, in the remainder of the paper we conversely examine how the evolution of a primordial non-vanishing Hamiltonian (due to any kind of global interaction) would change if the algebra of constraints were not the Dirac hypersurface deformation algebra~\cite{Dirac,DiracCanadian,Thiemann}, but were instead the algebra of HL theory. We limit ourselves to the projectable theory, so that the fluid equivalent to the non-vanishing Hamiltonian is a dust fluid~\cite{shinji,geoCDM}. The major difference can be expected from a central result in~\cite{geoCDM,nongeoCDM}: the Dirac hypersurface deformation algebra translates into local energy conservation (and Bianchi-like identities). Its violation in HL theory therefore implies local violations of energy conservation. These are to be distinguished for the violations induced by evolution~\cite{evol}, or global interactions~\cite{geoCDM}, which are global violations of energy conservation (absent in standard HL theory).

In this paper we explicitly compute the violations of the contracted Bianchi identities and corresponding violations of stress-energy conservation resulting from the breakdown of diffeormophism invariance, and so of the Dirac hypersurface deformation algebra. Furthermore we relate the divergence of these tensors to the various symmetry breaking couplings permitted within HL theory, each new allowed term translating into a source term for the divergence with a very specific tensorial structure. We do this in the simplest case  (Section~\ref{viol}) for clarity, then in full generality (Section~\ref{potentials} and Appendix~\ref{appendix}).

Beside its theoretical interest, this result has potentially very important implications towards our understanding of microscopic caustic avoidance, which is one of the most significant issues in many alternatives to particle dark matter scenarios.

\section{Projectable HL in a nutshell}\label{HL-review}

In the projectable Horava-Lifshitz (HL) gravity~\cite{HL}, the basic variables are 
\begin{equation}
\mbox{lapse}: N(t)\,, \ \mbox{shift}: N^i(t,x)\,, \ \mbox{3d metric}: h_{ij}(t,x)\,,
\end{equation}
out of which one can construct the spacetime metric at low energy as
\begin{equation}
 ds^4 = -N^2 dt^2 + h_{ij}(dx^i+N^idt)(dx^j+N^jdt)\,.
\end{equation}
The fundamental symmetry is the invariance under the foliation preserving diffeomorphism
\begin{equation}
 t \to t'(t)\,, \quad x^i \to {x'}^i(t,x)\,.
\end{equation}
In order to construct the kinetic terms for the spatial metric in a way that is compatible with the symmetry, one introduces the extrinsic curvature
\begin{equation}
    K_{ij} = \frac{1}{2N} (\dot{h}_{ij} - D_i N_j - D_j N_i)\,,
\end{equation}
where $D_i$ represents the spatial covariant derivative compatible with the spatial metric $h_{ij}$, an overdot represents the derivative with respect to $t$, and indices are lowered and raised using $h_{ij}$ and its inverse $h^{ij}$. The gravitational action is then written as
\begin{equation}
    S_{HL} = \frac{M_{Pl}^2}{2}\int{dt \: d^3x N\sqrt{h} \biggl [ (K_{ij}K^{ij}-\lambda K^2) -\mathcal{V}[h_{ij}] \biggl]}\,, \label{HL}
\end{equation}
where $M_{Pl}=1/\sqrt{8\pi G_N}$ is the reduced Planck mass, $h=\det h_{ij}$ is the determinant of the spatial metric, $K = K_{i}^{i}$ is the trace of the extrinsic curvature, and $\mathcal{V}[h_{ij}]$ is the Horava-Lifshitz potential, which contains spatial derivatives but not time derivatives.

The theory contains not only the usual tensor graviton but also a scalar graviton. In order to avoid the scalar graviton becoming a ghost, one needs to require the dimensionless coupling constant $\lambda$ to satisfy $\lambda<1/3$ or $\lambda>1$. In order for General Relativity (GR) or something akin to GR to be recovered at low energy, $\lambda$ has to run towards $1$ in the infrared (IR) under the renormalization group (RG) flow. Therefore, only the second range, i.e. 
\begin{equation}\label{eqn:lambda>1}
 \lambda>1\,,  
\end{equation}
remains viable. Phenomonologically, $\lambda$ has to approach $1$ sufficiently quickly towards the IR under the RG flow in order for the IR instability of the scalar graviton not to show up (see the more precise condition in (56) of \cite{shinji_review}). Under this condition, one can show that GR plus dark matter is recovered at low energy, at least for simple cases such as spherically symmetric, static configurations~\cite{shinji_review} and superhorizon nonlinear perturbations~\cite{Izumi:2011eh,Gumrukcuoglu:2011ef}. In this way the condensate of the scalar graviton behaves as cold dark matter at macroscopic scales, while it behaves quite differetly at microsccopic scales as we shall see in Sections~\ref{testtube} and \ref{potentials}.

\section{Time and time-dependence in HL gravity}\label{Sec:HL}

We begin from the Horava-Lifshitz action (\ref{HL}) and explore the implications of generalising the varying natural constants idea, as presented in ~\cite{evol}, ~\cite{Magueijo:2021pvq} and ~\cite{Magueijo:2021rpi}, in the context of the HL theory of gravity. Naturally, since the HL theory is built to be foliated, we assume some dynamical variable that are functions of the spacial leaves as a whole. In Mach's principle's spirit, we call them global variables. Additionally, we assume that the local HL action depends parametrically on these global variables. The possibilities for evolution in such theories are plenty: predominantly, the global variables provide relational clocks for the evolution of the local ones, as done in ~\cite{geoCDM}. However, another interesting possibility arises when the HL coupling constant $\lambda$ is used as a global variable for the evolution of another global variable, the cosmological constant, as we will see in what follows.

To enforce the evolution of the constants of Nature, the Henneaux-Teitelboim formulation~\cite{unimod} of unimodular gravity~\cite{unimod1, UnimodLee1, alan, daughton, sorkin1, sorkin2} is implemented, providing excellent definitions of physical relational times for the global variables such as $\Lambda$ and 4-volume or unimodular time ~\cite{Magueijo:2021pvq, Magueijo:2021rpi,Bombelli, UnimodLee2}. To set the stage for evolution up, we add to the HL base action the unimodular term such that:
\begin{equation}
    S_{HL} \rightarrow S = S_{HL} + S_U = S_{HL} + S_{NL} \label{nl},
\end{equation}
where, in the last step, we have relabelled the unimodular action $S_U$ as a non-local action $S_{NL}$ since the unimodular variables are assumed to be global. In fact, the total action can be written as:
\begin{align}
    S &= S_{HL} + \int{dt \: V_{\infty} \dot{\alpha} T_\alpha} \nonumber\\
    &= \int{dt \biggl[ \: \int_{\Sigma_t}{d^3 x \: (\dot{q}(x)p(x) - \mathcal{H}_E) + V_{\infty} \dot{\alpha}T_\alpha \biggl]}} \label{local_global_action},
\end{align}
where $V_{\infty} = \int_{\Sigma_t}{d^3 x}$ is the volume of $\Sigma_t$, $\mathcal{H}_E = N \mathcal{H} + N^{i} \mathcal{H}_{i}$ is the extended Hamiltonian density and the canonical pair $\{\alpha, T_\alpha \}$ consists of the global variables, while the pair $\{q(x), p(x)\}$ of the local ones from the HL action.
Importantly, both pairs are assumed to be intensive quantities, explaining the factor of $V_{\infty}$ in front of action \eqref{local_global_action}. Furthermore, the global variables are independent of position on $\Sigma_t$, being a property of the whole leaf. Conversely, the canonical pair of local variables characterize each point on the leaf.

It is now possible to quantify the effects of the global variables on the local ones studying the Hamiltonian structure of this theory. From the HL action we have derived the extended Hamiltonian density, which may depend on both local and global variables. To find the equations of motion, we define the total Hamiltonian as
\begin{equation}
    \mathbf{H} = \int_{\Sigma_t}{d^3 x \: \mathcal{H}_E(x)},
\end{equation}
and then, we proceed creating an intensive total Hamiltonian density:
\begin{equation}
    \mathcal{H}_T = \frac{\mathbf{H}}{V_{\infty}} = \frac{1}{V_{\infty}} \int_{\Sigma_t}{d^3 x \: \mathcal{H}_E(x)}.
\end{equation}
On the one hand, Hamilton's equations for the local variables can easily be derived from the extended Hamiltonian density, as done in ~\cite{geoCDM}. They imply that the evolution of local variables at a point depends locally on the derivative of the Hamiltonian at that point. On the other hand, the equation of motion for the global variables are:
\begin{align}
    \dot{\alpha} &= \{\alpha, \mathbf{H}\} = \frac{\partial \mathcal{H}_T}{\partial T_\alpha}\,,\\
    \dot{T}_\alpha &= \{T_\alpha, \mathbf{H}\} = -\frac{\partial \mathcal{H}_T}{\partial \alpha}\,. \label{time_eqn}
\end{align}
The evolution of global variables therefore depends on the Hamiltonian density over the whole leaf. As we will now see, the introduction of global variables dictating the evolution of local physics can be implemented in two ways, where the differentiation of $\bm{\beta}$-class constants and $\bm{\alpha}$-class one plays a crucial role.

The unimodular term in action \eqref{local_global_action} containing the global varibales can be generalised to account for multiple constants. In fact, as done in  ~\cite{evol, uni_BD}, we separate all the constants giving relational times (forming the $\bm{\alpha}$) from the ones that evolve under these time, forming a distinct set such that:
\begin{equation}
    \bm{\beta} = \bm{\beta} (\bm{T_{\alpha}}) \label{beta_alpha}.
\end{equation}
Therefore, in this setup for evolution, the global constants form the unimodular $\bm{\alpha}$ vector, while the local constants evolving under the canonical times $\bm{T_\alpha}$ are part of the $\bm{\beta}$ vector. With this global-time-driven local evolution, we consider two applications to the HL theory of gravity. Crucially, the presence of $\lambda$ in the kinetic term of the theory affects the minisuperspace (MSS) realisation of the theory.

To begin, we reduce action \eqref{HL} to MSS for an FLRW universe with metric:
\begin{equation}
    ds^2 = -dt^2 + a(t)^2 \biggl[ \: \frac{dr^2}{1-kr^2} + r^2 d\Omega^2 \biggl],
\end{equation}
where $a(t)$ is the scale factor, $k$ is the curvature constant taking values $k = \{-1, 0, 1\}$ and $ d\Omega^2 = (d\theta^2 + \sin{\theta}^2 d\phi^2)$ is the line element of the unit $2$-sphere, and taking care to choose the potential to be:
\begin{equation}
    -\mathcal{V}[h_{ij}] = 2 \Lambda + {}^{(3)}R. \label{gr_potential}
\end{equation}
This form ensures that the MSS action is equivalent to the GR one up to the coupling $\lambda$~\cite{Contillo:2013fua}, while different potentials for the UV (see ~\cite{shinji_review} for examples) would lead to different MSS realisations. Therefore, the purely gravity MSS HL action is:
\begin{equation}
    S_{HL} =  V_c \int{dt \: \alpha_2 \biggl[\dot{b}a^2 + Na \biggl[\biggl(\frac{3-9\lambda}{6}\biggl)b^2 + kc_g^2 \biggl] \biggl]}\,,
\end{equation}
where $V_c = \int{d^3 x}$ is the comoving volume, $b = \frac{\dot{a}}{N}$ is the MSS connection, $c_g^2$ is the gravity speed of light and we have defined the constants appearing in the Planck's mass as
\begin{equation}
    \alpha_2 = \frac{3 c_P^2}{8 \pi G_P}
\end{equation}
in light of ~\cite{evol} for varying constants theories. From this action, the MSS Hamiltonian can easily be obtained as:
\begin{equation}
     H = Na \biggl[-\alpha_2 (\alpha_{HL} b^2 + kc_g^2) + \alpha_3 \rho_\Lambda a^2 \biggl],
\end{equation}
where $\rho_\Lambda$ is the cosmological constant's energy density dual to the unimodular time $T_\Lambda$ and $\alpha_3 = \frac{G_M}{G_P}$. Also, $\rho_\Lambda$ comes from the matter HL action, given by the cosmological constant in potential \eqref{gr_potential}, following the same notation outlined in ~\cite{evol}. Importantly, we have defined the HL coupling constant as:
\begin{equation}
    \alpha_{HL} := \frac{3 - 9\lambda}{6},
\end{equation}
such that $\lambda$ can generate the global HL time under which some $\bm{\beta}$ evolves.

Setting $H \overset{!}{=} 0$, we have Hamilton's constraint,
\begin{equation}
    \alpha_{HL} b^2 + kc_g^2 = \frac{\alpha_3}{\alpha_2}\rho_\Lambda a^2.
\end{equation}
Furthermore, the equations of motion for $a$ and $b$ can be easily obtained from the Hamiltonian assuming $\dot{\alpha}_2 = \dot{\alpha}_3 = 0$ as:
\begin{align}
    \dot{a} &= Nb \alpha_{HL}\,, \\
    \dot{b} &= -\frac{\alpha_3}{2\alpha_2}(\rho_\Lambda + 3p_\Lambda) Na\,,
\end{align}
where $\rho_\Lambda$ and $p_\Lambda$ are the cosmological constant energy density and pressure respectively, related by the equation of state $p_\Lambda = -\rho_\Lambda$. Finally, dotting the Hamiltonian constraint we arrive at a general conservation equation:
\begin{equation}
    \dot{\rho}_\Lambda + 3\frac{\dot{a}}{a}(\rho_\Lambda + p_\Lambda) = \frac{k \alpha_2}{\alpha_3 a^2}\frac{dc_g^2}{dt} + \frac{\dot{\alpha}_{HL}}{\alpha_{HL}^2}\frac{\alpha_2}{\alpha_3}\frac{\dot{a}^2}{N^2 a^2}. \label{cons}
\end{equation}
We will now explore two different scenarios. In the first one, $\lambda$ will give the HL time driving the evolution of the  gravity speed of light $c_g^2$. In the second case, it will be $\lambda$ to evolve according to the unimodular time $T_{\Lambda}$ in an interesting hybrid between unimodular and RG evolution.

\subsection{A simple model: $c_g^2 = c_g^2 (T_{\alpha_{HL}})$} \label{c_evol}

We consider a varying speed of light theory where $c_g^2$ depends on the HL time given by $\alpha_{HL}$. In this case, beyond the usual $a$ and $b$ equations of motion, we also have:
\begin{align}
    \dot{\alpha}_{HL} &= \frac{\partial H}{\partial T_{\alpha_{HL}}} = - Na\alpha_2 k \frac{\partial c_g^2}{\partial T_{\alpha_{HL}}}, \label{alpha_hl} \\
    \dot{T}_{\alpha_{HL}} &= -\frac{\partial H}{\partial \alpha_{HL}} = Na\alpha_2 b^2.
\end{align}
Also, since $c_g^2$ depends on $T_{\alpha_{HL}}$, we can write:
\begin{equation}
    \frac{dc_g^2}{dt} = \frac{\partial c_g^2}{\partial T_{\alpha_{HL}}}\dot{T}_{\alpha_{HL}},
\end{equation}
leading to
\begin{equation}
    \dot{\rho}_\Lambda + 3\frac{\dot{a}}{a}(\rho_\Lambda + p_\Lambda) = \frac{\alpha_2^2 k N}{\alpha_3 a}\frac{\partial c_g^2}{\partial T_{\alpha_{HL}}}b^2 - \frac{\alpha_2^2 k}{\alpha_3 a N}\frac{\partial c_g^2}{\partial T_{\alpha_{HL}}}\frac{\dot{a}^2}{\alpha_{HL}^2}.
\end{equation}
The equation above leads to full energy conservation once the $\dot{a}$ equation is enforced, as expected. More interestingly, equation \eqref{alpha_hl} can be used to derive a relation between $\alpha_{HL}$ and $c_g^2$:
\begin{equation}
    \dot{c}_g^2 = -\frac{b^2}{k}\dot{\alpha}_{HL} \label{varying_c}.
\end{equation}
This confirms results previously obtained in ~\cite{uni_BD} and opens new prospectives on the idea of varying speed of light theories ~\cite{vsl}. In fact, equation \eqref{varying_c} has the correct sign for a decreasing $c_g^2$ from the early universe till today for a $k = 1$ FLRW universe. We speculate that perhaps a theory of gravity evolving in different energy regimes according to $\lambda$ might, in the future, shed light on the mechanisms behind a varying speed of light in the early universe.

\subsection{Unimodular evolution and RG flow: $\lambda = \lambda (T_{\Lambda})$}

We now explore the other possibility for evolution: $\lambda$ evolving under unimodular time $T_\Lambda$. In this case, we are using a global time variable to drive the evolution of another global variable, since $\lambda$ determines the energy limit of the HL theory on the entire leaf. To complicate the picture, $\lambda$ is also subject to RG flow since it determines the energy scale $\mu$ of the theory, as pointed out in ~\cite{HL, shinji_review}. The aim is to create an hybrid between these two mode of evolution, such that unimodular evolution can mimic RG running with energy.
To begin, we briefly review unimodular evolution and RG flow running, highlighting the link between the two. On the one hand, unimodular evolution is given by the canonical variable $T_\alpha$ appearing in action \eqref{local_global_action}, once we postulate \eqref{beta_alpha}, such that $\lambda = \lambda(T_\Lambda)$. On the other hand, RG flow is regulated by the beta function ~\cite{RGflow}, which relates $\lambda$ and $\mu$ as
\begin{equation}
    \beta (\lambda) = \frac{d \lambda}{d \ln{\mu}} \label{beta_func}.
\end{equation}
Importantly, the specific form of $\beta$ will give different fixed points of the RG flow, affecting the allowed values of $z$, the HL critical exponent. Given this expression and unimodular evolution, we can link $T_\Lambda$ expressed in term of $\mu$ with $\beta$, finding a unimodular RG flow evolution.

Using MSS variables as in ~\cite{evol} and equation \eqref{time_eqn}, the unimodular time is defined as: 
\begin{equation}
    T_{\Lambda} = \int{d^4 x \: a^3} \label{uni_time}.
\end{equation}
Then, the scale factor for an FLRW radiation-dominated universe is used, since this model gives the early stages of the universe's evolution, when also RG flow becomes important. This is given by 
\begin{equation}
    a = a_0 \biggl(\frac{t}{t_0}\biggl)^{\frac{1}{2}} \label{scale_rad},
\end{equation}
where $t$ is cosmic time and $a_0$ and $t_0$ are initial conditions. Furthermore, the scale factor can be related with the temperature $T$ of the universe with
\begin{equation}
    a = \frac{a_0 T_0}{T} = \frac{a_0 \mu_0}{\mu} \label{scale_temp},
\end{equation}
where, in the last step, we have assumed that temperature is related to energy scale: $T \sim E \sim \mu$, as it is natural if scalar or tensor gravitons are thermalized. Plugging equation \eqref{scale_rad} into equation \eqref{uni_time}, we perform the integration obtaining:
\begin{equation}
    T_{\Lambda} (\mu) = \frac{2}{5}t_0 \: a_0^3 \: \mu_0^5 \frac{1}{\mu^5} \label{int_time}, 
\end{equation}
where we have expressed the scale factor using \eqref{scale_temp} and we have dropped any integration constant for simplicity. Therefore, we have an expression for the dependence of unimodular time on energy scale. In principle, if we can relate the energy scale $\mu$ to $\lambda$ under RG flow, we can then connect $\lambda$ with unimodular time. 

Proceeding, we construct a polynomial $\beta$ function for $\lambda$, such that equation \eqref{beta_func} becomes:
\begin{equation}
    \beta (\lambda) = \frac{d \lambda}{d \ln{\mu}} = n (\lambda - 1),
\end{equation}
where $n$ is the order of polynomial. Solving for $\lambda$ as a function of energy scale, we obtain a polynomial running,
\begin{equation}
    \lambda(\mu) = 1 + \biggl(\frac{\mu}{\mu_0} \biggl)^n, \label{rg_fun}
\end{equation}
where $\mu_0$ is a positive constant. Finally, we can invert this equation expressing $\mu$ in terms of $\lambda$, plug the result into equation \eqref{int_time} to then get:
\begin{equation}
    \lambda(T_\Lambda) = 1 + \biggl[\biggl(\frac{2}{5}t_0 \biggl)^{\frac{1}{5}}a_0^{\frac{3}{5}}T_\Lambda^{-\frac{1}{5}}\biggl]^n. \label{rg_uni_hybrid}
\end{equation}
This equation describes the dependence of the HL coupling on unimodular time in a RG flow fashion, since we have designed the function from equation \eqref{beta_func}. Beside the theoretical interest, this new hybrid could lead to potentially relevant results when applied to an FLRW universe. As pointed out in ~\cite{shinji_review}, in HL theory the local Hamiltonian constraint is lost, so that we can only consider the dynamical equation:
\begin{equation}
    \frac{3(3\lambda-1)}{2}H^2 = 8\pi G \biggl[\rho +\frac{C(t)}{a^3} \biggl],
\end{equation}
where $H$ is the Hubble parameter and $C(t)$ is an integration constant (for more details consult ~\cite{shinji_review}). This modification implies that the Hubble expansion rate is smaller if large values of $\lambda$ are considered, deviating from the standard GR result. Equation \eqref{rg_uni_hybrid} could then be used to improve the FLRW singularity by making $\lambda \rightarrow \infty$ in the RG flow. Of course, this possibility remains to be largely explored in future work, potentially considering different RG running other than \eqref{rg_fun}.

\section{Dark Matter with violations of energy conservation}\label{viol}
HL is subject to a {\it global} Hamiltonian constraint, but the {\it local} Hamiltonian constraint is generally violated~\cite{shinji}. This is identical to what happens to General Relativity subject to time-dependence or global interactions~\cite{geoCDM}. This last case motivates the introduction of the non-local action in equations \eqref{nl} and \eqref{local_global_action}, as they describe the effects of global interactions on the Hamiltonian structure of GR. However, unlike in the GR case, for HL we do not assume any non-local interactions a priori, since the algebra is already modified by the different theory. As long as the preferred foliation is geodesic, in both cases the non-vanishing Hamiltonian can be reinterpreted as an effective dust fluid in a theory satisfying a local Hamiltonian constraint.  That is, one can redefine the non-vanishing Hamiltonian ${\cal H}$ (which contains gravity and ``normal matter'') as a new vanishing Hamiltonian $\bar {\cal H}$ at the expense of introducing a new dark matter component ${\cal H}_m=-{\cal H}$, so that:
\begin{equation}
    {\cal H}\neq 0 \rightarrow \bar {\cal H}={\cal H}+{\cal H}_m\approx 0\,. \label{h_mh_g}
\end{equation}
The regained Hamiltonian constraint can then be 
enforced by allowing 
\begin{equation}\label{Nofx}
    N(t)\rightarrow N(t,x)
\end{equation}
in the variations
(even though $N=N(t)$ on-shell, required by the geodesic nature of $\Sigma_t$). If $\Sigma_t$ is geodesic the local momentum constraint is preserved, ${\cal H}_i=0$, leading to the stress energy tensor of a dust fluid with 3 out of its 4 degrees of freedom frozen in~\cite{geoCDM} (or 3 first class constraints imposed).

However, there is a major difference between the effective dust associated with violations of the local Hamiltonian constraint in GR or any theory which recovers 4D diffeomorphism invariance when global interactions/time-dependence switch off, and in HL. In HL, the dust fluid does not satisfy stress-energy conservation. This crucial difference results from the fact that 
HL does not satisfy the Dirac Hypersurface deformation algebra~\cite{Dirac,DiracCanadian,Thiemann}. Note that the new matter component starts off life as a ``non-real'' matter component, since its evolution is fundamentally dictated by the all the other degrees of freedom effect upon the original non-vanishing Hamiltonian:
\begin{equation}
       \dot{\mathcal{H}} = \{\mathcal{H}, \mathbf{H} \}\,,
\end{equation}
where
\begin{equation}
     \mathbf{H}=\int d^3x\ (N{\cal H} +N^i {\cal H}_i).
\end{equation}
The dynamical evolution of this dust is therefore dictated by the algebra of constraints. 

For action \eqref{HL} we have Hamiltonian and momentum densities:
\begin{align}
     \mathcal{H} &= \frac{2}{M_{Pl}^2\sqrt{h}} \biggl[\pi_{ij}\pi^{ij} - \biggl ( \frac{\lambda}{3\lambda -1}\biggl)\pi^2\biggl] +\sqrt{h} \: \mathcal{V}[h_{ij}]\,,\\
     \mathcal{H}_i &= -2h_{ij} D_k \pi^{jk}\,,
\end{align}
leading to the algebra~\footnote{From the point of view of~\cite{geoCDM,nongeoCDM} where global variables are introduced, these would be the local Poisson brackets, that is the brackets containing only the local variables.}:
\begin{align}
    \{H_i(N^i), H_j(M^j)\}&= H_i([N,M]^i)\,,\label{smearhihi}\\
    \{H_i(N^i), H(N)\}&= H(N^i\partial _iN)\,,\label{smearhih0}\\
    \{ H(N), H(M)\}&= C_i(h^{ij}(N\partial_j M- M\partial_j N))\,, \label{smearh0h0}
\end{align}
where we used the smeared form of the constraints, $H(N)=\int d^3x N{\cal H}$, $H_i(N^i)=\int d^3 x N^i {\cal H}_i$
and $C_i(N^i)=\int d^3 x N^i {\cal C}_i$
with:
\begin{equation}
    {\cal C}_i = {\cal H}_i + 2 \hat{\lambda}D_i \pi\equiv
{\cal H}_i + 2 \frac{\lambda - 1}{3\lambda - 1}D_i \pi. 
\end{equation}
In this Section we chose potential \eqref{gr_potential}, but later we will generalize this result.

While for other potentials the reader is referred to~\cite{HLDiracal} (which reduces to our result for our choice of potential). Note that to comply with the conventions in use in the HL community we use the ``+'' sign convention for the extrinsic curvature (i.e. $\dot h_{ij}\rightarrow K_{\ij}$; see~\cite{HL}); this is the opposite convention used in~\cite{geoCDM,nongeoCDM,ADMReview}, following for example from the classic~\cite{MTW}. This trivial sign convention is of course physically irrelevant, but keeping consistency has important physical consequences, as we will see. 
A non-zero ``leftover'' local Hamiltonian  then has to be propagated just as in~\cite{geoCDM}, using the algebra \eqref{smearhihi}-\eqref{smearh0h0}. This leads to:
\begin{equation}
    \dot{\mathcal{H}} = \{\mathcal{H}, \mathbf{H} \} = \partial_i (N^i \mathcal{H}) + 2\hat{\lambda}N D^2 \pi\,,\label{main}
\end{equation}
where we have used the fact that the lapse function is projectable and that the momentum constraint is conserved in the preferred foliation $\Sigma_t$:
\begin{equation}
    \dot{\mathcal{H}}_i = \{\mathcal{H}_i, \mathbf{H} \} = 0 \Rightarrow \mathcal{H}_i = 0. 
\end{equation}
Since ${\cal H}_m=-{\cal H}$, 
the evolution of the new dark matter Hamiltonian therefore is given by:
\begin{equation}
    \dot{\mathcal{H}}_m = \partial_i (N^i \mathcal{H}_m) - 2\hat{\lambda}N D^2 \pi \label{main}
\end{equation}
with ${\cal H}^m_i=0$.

As announced,
we can trade the non-vanishing Hamiltonian by a new effective fluid with the same Hamiltonian and momentum and a local Hamiltonian constraint. On $\Sigma_t$ the only non-vanishing component of $T^m_{\mu\nu}$ for this fluid is:
\begin{equation}
    T^m_{00} = -\frac{2}{N \sqrt{h}}\frac{\delta S_m}{\delta g^{00}} = \frac{N^2}{\sqrt{h}}{\cal H}_m 
\end{equation}
(given that it only has a Hamiltonian, but no momentum or dependence on $h_{ij}$, i.e. stresses). 
Thus, this effective matter component takes the form of a dust fluid:
\begin{equation}
    T^m_{\mu\nu}=\rho_m u_\mu u_\nu \label{stress_tensor}
\end{equation}
with velocity vector aligned with the normal to $\Sigma_t$ (i.e. $u_\mu=n_\mu=(-N,0)$) and:
\begin{equation}
    \rho_m = \frac{{\cal H}_m}{\sqrt{h}}.
\end{equation}

Therefore, it would appear that just as in~\cite{geoCDM}, the legacy effect of time-variation is a residual Hamiltonian, equivalent to a dust fluid with rest frame aligned with the geodesic preferred $\Sigma_t$. However, a crucial difference arises when we verify its conservation equation. The fact that $\Sigma_t$ is geodesic ensures that the projected components of the divergence vanish:
\begin{equation}
    n^\mu\nabla_\mu n^\nu=0\implies h_{\alpha\nu}\nabla_\nu T^{\mu \nu}_m =0\,.
\end{equation}
However the ``time'' component leads to:
\begin{equation}
   - n_\mu \nabla_\nu T^{\mu \nu}_m = \nabla_\mu (\rho_m n^\mu) 
   =\frac{1}{\sqrt{-g}}\biggl[\dot{\cal H}_m-\partial_i(N^i{\cal H}_m)\biggl]
   \label{relat}
\end{equation}
so that \eqref{main} leads to a violation:
\begin{equation}
    - n_\mu \nabla_\nu T^{\mu \nu}_m = -2 c_g^2 \frac{\hat{\lambda}}{\sqrt{h}}D^2 \pi
\end{equation}
the culprit, understandably, being the new term modifying the Dirac Hypersurface Deformation algebra. Using 
\begin{equation}
    \pi = \frac{M_{Pl}^2}{2}\sqrt{h}(1-3\lambda)K
\end{equation}
this can also be written as:
\begin{equation}
     -n_\mu \nabla_\nu T^{\mu \nu}_m = c_g^2 M_{Pl}^2 (\lambda-1) D^2 K \label{main_res}
\end{equation}
so the violations depend directly on the deviation of $\lambda$ from its GR value, as well as from the spatial Laplacian of the trace of the extrinsic curvature of the preferred foliation. We note that this effective dark matter component is exactly the same one found from equation $(4.13)$ of ~\cite{shinji}. Furthermore, while the source term was only shown to be proportional to $(\lambda-1)$ in ~\cite{shinji}, here the full dependency on the extrinsic curvature was derived explicitly. Finally, we remark that for FLRW the source term of \eqref{main_res} vanishes, making the effective dark matter evolution equivalent to standard cold dark matter.

\section{A test tube:  the Schwarzschild solution in Lemaitre coordinates}\label{testtube}

The source term found in the previous section for the projectable HL theory is expected to play an important role in the early universe and for the short distance behavior of gravity. At long distances and at low energies compared with the quantum gravity scale, $(\lambda-1)$ is required to be extremely small by  phenomenological constraints (see (56) of \cite{shinji_review}), implying that one can completely ignore the source term (the higher curvature corrections ignored in the previous section are also obviously unimportant) and that the effective dark matter evolution is equivalent to that of the standard cold dark matter including self-gravity. Needless to say, this means that the effective dark matter component in the projectable HL theory can explain a variety of observational data, including galaxy rotation curves, at least to the same degree as the standard cold dark matter. This situation is similar to mimetic gravity (see e.g.~\cite{mimetic}).

However, unlike mimetic gravity, at short distances and at high energies in the UV regime, the behavior can be very different from that of the standard cold dark matter because of the source term. The parameter $\lambda$ (with $\lambda>1$) should deviate significantly from $1$ due to the RG running, and the higher curvature terms become also important, altogether making the source term active. The form of the source terms found in the previous section and in the next section may then explain how, unlike with mimetic gravity, one can avoid caustics at short distances~\cite{Mukohyama:2009tp}, one of the most important issues in any gravity theory containing an effective dark matter component. 

As a simple illustration of the role of the source term and the energy conservation violation at microscopic scales, we investigate here the behavior of the effective energy density of the dark matter in the Schwarzschild background without taking into account the backreaction. Lemaitre coordinates, a generalisation of curvature coordinates, are used. Importantly, these coordinates can be used because we are working with the projectable version of HL theory, implying $n^\mu = (1, \mathbf{0})$. 

The Lemaitre metric, as presented in ~\cite{stephani, Enqvist:2007vb, Goncalves:2001rv}, takes the form:
\begin{equation}
    ds^2 = -dt^2 + \frac{(A^\prime)^2}{1-k(r)}dr^2 + A^2 d\Omega^2\,, \label{S_L_metric}
\end{equation}
where $A$ is the areal radius and prime denotes the derivative with respect to the coordinate $r$.
More evocatively, equation \eqref{main_res} can be written in terms of $\rho_m$ using the dust fluid stress energy momentum \eqref{stress_tensor}, such that:
\begin{equation}
    -n_\mu \nabla_\nu T^{\mu \nu}_m = \dot{\rho}_m + \rho_m \nabla_\mu n^{\mu}, \label{intermidiate}
\end{equation}
where we have used that the fluid's velocity is normal to $\Sigma_t$ and that in Lemaitre coordinates $n^\mu = (1, \mathbf{0})$. Furthermore, using metric \eqref{S_L_metric}, the extrinsic curvature scalar takes the form:
\begin{equation}
     K = \nabla_\mu n^\mu = 2\frac{\dot{A}}{A} + \frac{\dot{A}^\prime}{A^\prime}.
\end{equation}
Therefore, combining these results, \eqref{main_res} is brought into the form of a general conservation equation with an energy-violating source term given by the extrinsic curvature:
\begin{equation}
    \dot{\rho}_m + \rho_m \biggl[2\frac{\dot{A}}{A} + \frac{\dot{A}^\prime}{A^\prime}\biggl] = c_g^2 M_{Pl}^2 (\lambda-1) D^2 K \label{general_non_cons}.
\end{equation}

Proceeding, we choose the $k(r) = 0$ solution for $A$, giving, in Lemaitre coordinates:
\begin{equation}
    A = \biggl (\frac{9}{4}F\biggl)^\frac{1}{3} |t_0-t|^\frac{2}{3}, \label{A_K_zero}
\end{equation}
where $t_0 = r-r_0$ is a free function related to the coordinate radius $r$ and $F(r)$ is an independent function, being $F = 2 r_s$ for a Schwarzshild spacetime. There are actually two branches: the contracting one with $t<t_0$ and the expanding one with $t>t_0$. The covariant derivative's definition for this solution becomes:
\begin{equation}
    D^2 K = \frac{1}{A^2 A^{\prime}} \: \partial_r \biggl[\frac{A^2}{A^{\prime}}\: \partial_r K\biggl],
\end{equation}
given that $F$ can be treated as a constant and therefore $\dot{A} = A^{\prime}$. Consequently, we plug this result into equation \eqref{general_non_cons}, obtaining the ordinary differential equation:
\begin{equation}
    \frac{d \rho_m}{dt} = -\frac{\rho_m}{|t - t_0|} + \frac{3}{4}c_g^2 M_{Pl}^2 (\lambda-1)\biggl (\frac{9}{4}F\biggl)^{-\frac{2}{3}} |t_0-t|^{-\frac{7}{3}}\,.
\end{equation}
This differential equation can be separated in its homogeneous and inhomogeneous parts and solved separately. On the one hand, the homogenous part integrates giving the free solution:
\begin{equation}\label{eqn:homogeneous-sol}
    \rho_m = \frac{\rho_0}{|t_0 - t|},
\end{equation}
where $\rho_0$ is an arbitrary constant. On the other hand, the inhomogeneous part can be integrated following standard techniques, giving~\footnote{The driven solution, expressed in terms of the areal radius $A$, takes the form: $\rho_m = - (9/4) c_g^2 M_{Pl}^2(\lambda-1)/A^2$. This appears tantalizing as it is precisely the correct profile required for flattening the rotation curves of galaxies. In fact, the rotational velocity of the outer-most objects in galaxies are observed to be constant ~\cite{vera_rubin}. To keep the velocity constant, the mass enclosed in the galaxy must scale linearly with the radius, which is only possible when the density profile is the one found here. However, we know that this cannot be a realistic explanation for such an observation since $\rho_m$ is negative. Furthermore, we assumed that the dark matter had negligible active gravity in this picture, and contented ourselves to study its passive gravity in the field of a more massive central black hole.}:
\begin{equation}\label{eqn:driven-sol}
    \rho_m = -\frac{9}{4} c_g^2 M_{Pl}^2(\lambda-1) \biggl(\frac{9}{4}F \biggl)^{-\frac{2}{3}} |t_0 -t|^{-\frac{4}{3}}. 
\end{equation}
The general solution is a sum of \eqref{eqn:homogeneous-sol} and \eqref{eqn:driven-sol}, with $\rho_0$ being an integration constant. Comparing the (negative) powers of $|t_0-t|$ in \eqref{eqn:homogeneous-sol} and \eqref{eqn:driven-sol}, we see that the driven solution \eqref{eqn:driven-sol} dominates over the homogeneous solution \eqref{eqn:homogeneous-sol} at late time and thus is an attractor of the system if and only if the contracting branch with $t<t_0$ (instead of the expanding one with $t>t_0$) is chosen. The sign of the driven solution \eqref{eqn:driven-sol} is negative (see \eqref{eqn:lambda>1}). Considering the contracting branch, the driven solution \eqref{eqn:driven-sol} is the attractor of the system and the source term would provide the necessary negative energy to avoid caustics in the projectable HL gravity~\cite{Mukohyama:2009tp}. 

We should admit that this treatment is potentially too idealised. A passive dark matter configuration with negative energy density around a central black hole has been considered; a more realistic scenario with self-gravity and matter fields included would make the setup more realistic. This is, however, beyond the scope of this paper. In our treatment we are considering the negative mass of the halo $M_{h}$ to be much smaller then the Schwarzshild mass $M_{s}$, such that $|M_{h}| << M_{s}$, so that self-gravity can be ignored. Our passive gravity treatment is justified only under this condition: we defer the active scenario to future work ~\cite{active_dm}. 

We do take the first steps in this direction here.
Once the self-gravity, or the backreaction, is taken into account, the spatial curvature of the constant time hypersurface is expected to grow and thus higher curvature corrections should also be taken into account. For this reason, in the next section we shall study those higher curvature corrections to the source term.

\section{higher curvature corrections: a generalisation} \label{potentials}

The result obtained in the previous section is extremely simplified and depends on several choices made at the beginning of the derivation. A crucial one is the potential \eqref{gr_potential}, which constrains the theory to the GR limit if $\lambda\rightarrow 1$. This condition can be relaxed adding terms going  beyond the usual $z=1$ scaling in the IR to the HL potential. Following~\cite{shinji_review}, action \eqref{HL} can be decomposed into several terms:
\begin{equation}
    S_{HL} = S_{K.T.} + S_{z=1} + S_{z=0} + S_{z > 1},
\end{equation}
where the first term is the usual kinetic contribution, while the second and third ones simply give the GR potential used in \eqref{gr_potential}. Otherwise, $S_{z > 1}$ contains the higher curvature corrections to the GR potential:
\begin{align}
    S_{z>1} &= \frac{M_{Pl}^2}{2} \int{dt \: d^3x \sqrt{h}N \: [ (c_1 D_i R_{jk}D^{i}R^{jk} + c_2 D_i R D^{i}R} \nonumber \\
    &+c_3 R^{j}_{i}R^{k}_{j}R^{i}_{k} + c_4 R R^{j}_{i}R^{i}_{j} + c_5 R^3) \nonumber \\
    &+ (c_6 R^{j}_{i} R^{i}_{j} + c_7 R^2)] ,
\end{align}
where the first bracket contains the terms given by the UV contributions with $z > 2$, while the second one determines the deformations in the IR given by $z = 2$. Including these higher curvature terms, the potential changes, affecting the Dirac's algebra \eqref{smearh0h0} and ultimately generalising equation \eqref{main_res} to:
\begin{equation}
    -n_{\mu}\nabla_{\nu}T^{\mu \nu}_m =  n^{\mu}\partial_{\mu} \rho_m + K \rho_m = c_g^2 s_0 + \sum_{i=1}^{7}c_i s_i\,, \label{general_non}
\end{equation}
where $c_i$ are constants and $s_i$ are the various energy violating contributions. Specifically,\begin{equation}
 s_0 = (\lambda - 1)M_{Pl}^2 D^2 K \,, 
\end{equation}
is the first order term already found in \eqref{main_res}, while the other lead higher order contributions. As an appetiser, we present the $s_7, s_6$ and $s_5$ terms, leaving the other terms to Appendix~\ref{appendix} as the main course. These will be the foundations for future work ~\cite{active_dm}, where we will investigate the active and self gravity of our dark matter.
Hence, the three source terms read:
\begin{align}
    s_7 &= 4 \left[ (\lambda - 1) R D^2 K + K D^2 R - K^{ij}D_iD _jR \right]\,, \label{s_7} \\ 
    s_6 &= KD^2R + 2K^{ij}D^2R_{ij} - 2R_ {ij}D^2K^{ij} \nonumber \label{s_6} \\
 &  + 2(2\lambda -1)R^{ij}D_iD _jK - 2K^{ij}D_iD _jR\,, \\ 
    s_5 &= 6[(\lambda-1)R^2 D^2K+2KRD^2R \nonumber \\
    &- 2RK^{ij}D_i D_j R + 2KD^i R D_i R \nonumber \\
    &- 2K^{ij}D_i R D_j R], \label{s_5}
\end{align}
where we notice that there are various combinations of different powers of extrinsic and intrinsic curvatures as well as spatial covariant derivatives. Importantly, we see that energy density is not conserved either if $\lambda \neq 1$ or if higher curvature corrections are included in the theory, confirming previous results obtained in ~\cite{shinji}.

Turning this result to Lemaitre coordinates with $k(r)=0$, where $R_{ij}=0$ and $R=0$, all the source terms vanish except for $s_0$. This justifies the treatment in the previous section, as far as the gravitational backreaction is negligible.

However, once the negative amplitude \eqref{eqn:driven-sol} of the driven (and attractor) solution becomes sufficiently large during the contraction, the backreaction cannot be ignored. It is then expected that spatial curvature of the constant time hypersurfaces should deviate from zero and the source terms other than $s_0$ should become active. We expect that they all together help avoiding the caustic formation~\cite{Mukohyama:2009tp}.

\section{Conclusion}
In this paper we examined a little-known aspect to HL theory: that at high energies and short distances it implies violations of energy conservation as a result of the breakdown of Bianchi identities and 4D diffeomorphism invariance. The matter is well-known in theories with global interactions, such as those with time evolution in the laws of physics~\cite{evol,geoCDM,nongeoCDM}. In such theories the free evolution of the left-over Hamiltonian (and momentum, if the preferred foliation is not geodesic) maps them into an effective fluid carrying the non-vanishing Hamiltonian (and momentum, in the non-projectable theory). It was proved in~\cite{geoCDM} (in full generality in~\cite{nongeoCDM}) that restoration of the algebra of constraints of General Relativity is equivalent to stress-energy conservation for this effective fluid. If the algebra of HL persists we therefore have to face violations of energy conservation. In the simplest model these are proportional to the Laplacian of the extrinsic curvature of the preferred foliation and to the deviation of the parameter $\lambda$ from $1$. We evaluated the source term for the full class of projectable HL models in this paper.

The source term found in this paper for the full class of the projectable HL theory is expected to play important roles in the early universe and for the short distance behavior of gravity. At long distances and at low energies compared with the quantum gravity scale, the higher curvature corrections found in section~\ref{potentials} are obviously unimportant and $(\lambda-1)$ is required to be extremely small by the phenomenological constraint (see (56) of \cite{shinji_review}), meaning that one can completely ignore the leading source term shown in (\ref{main_res}) as well and that the effective dark matter evolution is equivalent to that of the standard cold dark matter including self-gravity. Needless to say, this means that the effective dark matter component in the projectable HL theory can explain a variety of observational data, including galaxy rotation curves, at least to the same degree as the standard cold dark matter. However, at short distances and at high energies in the UV regime, the behavior can be very different from that of the standard cold dark matter because of the source term. The $\lambda$ should deviate significantly from $1$ due to the RG running and the higher curvature terms are also important, making the source term active. For example, the form of the source term found in the present paper may explain how one can avoid caustics at short distances~\cite{Mukohyama:2009tp}, one of the most important issues in any gravity theory containing an effective dark matter component. Furthermore, considering contracting geodesics, the driven solution \eqref{eqn:driven-sol} is the attractor of the system. Therefore, the source term would provide the necessary negative energy to avoid caustics in the projectable HL gravity.

Finally, as a generalization of the source term shown in (\ref{main_res}) and in section~\ref{potentials}, one can easily include non-conservation of  the ordinary matter stress-energy tensor, as shown in (4.13) of \cite{shinji}. Such non-conservation is induced by higher spatial derivative terms present in the ordinary matter action, embedded in HL theory, and becomes important in the early universe, leading to interactions between the effective dark matter component and ordinary matter. All three terms on the right hand side of (4.13) of \cite{shinji} are important only at high energy. The first term was already explicit in \cite{shinji} while the second and third terms were made explicit in the present paper for the first time. Therefore, the explicit expression of the source term found in the present paper is expected to be useful for exploration of cosmological implications of such interactions as well.

\section*{Acknowledgments}
We thank Chris Isham and Raymond Isichei for help with this paper, Lorenzo Signore for insightful discussions and Alexander Cerato for great inspiration.
This work was partly supported by the STFC Consolidated Grants ST/T000791/1 and ST/X00575/1 (JM), JSPS KAKENHI Grant No.\ JP24K07017 (SM), and World Premier International Research Center Initiative (WPI), MEXT, Japan (SM).

\appendix
\section{Remaining source terms} \label{appendix}

As promised in section \ref{potentials}, we include here for completeness the remaining energy violating terms. Additionally, we report the extrinsic curvature computed for solution \eqref{A_K_zero}, along with the two most frequent derivative terms, as well as equation \eqref{general_non} including the $s_7$, $s_6$ and $s_5$ terms. The remaining $s_i$ terms are:
\begin{widetext}

\begin{align}
 s_1  &= -2 D_i R_{jk} D^i D^2 K^{jk} + 
 2 (2\lambda - 1) D_i R_{jk} D^i D^j D^k K - 4 R_i^{\ l} D_j R_{lk} D^k K^{ij} + 8 R_i^{\ l} D_j R_{lk} D^j K^{ik} \nonumber\\
&  + 4 R_i^{\ l} D_j R_{lk} D^i K^{jk} - 4 R_{jk} D_i R D^i K^{jk} - 2 R D_i R_{jk} D^i K^{jk} - 4 R^{ij} D_k R_{ij} D^i K + 2 R D_i R D^i K \nonumber\\
&  - 2 K^{ij} D^2 D^2 R_{ij} + 2 K^{ij} D^2 D_i D_j R - K D^2 D^2 R + 4 K^{ij} R^{kl} D_k D_l R_{ij} + 12 K^{ij} R^{kl} D_i D_l R_{kj} \nonumber\\
&  - 8 K^{ij} R^{kl} D_i D_j R_{kl} - 6 R K^{ij} D^2 R_{ij} - 2 K R^{ij} D^2 R_{ij} + 16 K^{il} R_l^{\ j} D^2 R_{ij} - 14 K^{il} R_l^{\ j} D_i D_j R + 4 R K^{ij} D_i D_j R \nonumber\\
&  - 2 K^{ij} R_{ij} D^2 R + K R D^2 R - 4 K^{ij} D_i R_{kl} D_j R^{kl} - 4 K^{ij} D_i R_{kl} D^k R_j^{\ l} + 24 K^{ij} D_k R_{il} D^k R^l_{\ j} \nonumber\\
&   - 8 K^{ij} D_k R_{il} D^l R^k_{\ j} - 6 K D_k R_{ij} D^k R^{ij} - 12 K^{ij} D^k R D_k R_{ij} + 6 K^{ij} D^k R D_i R_{kj} + 3 K D_i R D^i R \nonumber\\
&    - 36 K^{ij} R_{ik} R^{kl} R_{lj} + 30 R K^{ij} R_{ik} R^k_{\ j} + 12 R^{ij} R_{ij} K^{kl} R_{kl} - 12 R^2 K^{ij} R_{ij} - 12 K R_{ij} R^{jk} R_k^{\ i} + 6 K R R^{ij} R_{ij}\,, \\
 s_2 & = -R_{ij} D_k R D^k K_{ij} + (\lambda - 1) D^i R D_i D^2 K - K^{ij} D^k R D_k R_{ij} + K^{ij} D_i D_j D^2 R - K D^2 D^2 R - D^2 R K^{ij} R_{ij}\,, \\
 s_3 & = -R_{ik} R^{k}_{\ j} D^2 K^{ij} + (2\lambda - 1) R^{ik} R_k^{\ j} D_i D_j K - 2 K^{ik} R^{jl} D_i D_j R_{kl} + 2 K^{il} R_l^{\ j} D^2 R_{ij} - 2 K^{ij} D_i R^{kl} D_k R_{jl}  \nonumber \\
 &+  2 K^{ij} D^k R_i^{\ l} D_k R_{jl} + K D_i R_{jk} D^j R^{ik} - K^{ij} D^k R D_i R_{jk} - K^{ik} R_{kj} D^j D_i R + K R^{ij} D_j D_i R  + \frac{1}{4} K D^i R D_i R \nonumber\\
&   + K \left(\frac{1}{2} R^3 - \frac{5}{2} R R^{ij} R_{ij} + 3 R^{ij} R_{jk} R^k_{\ i}\right)\,, \\
 s_4 & = -2 R R_{ij} D^2 K^{ij} + 2 (2\lambda - 1) R R^{ij} D_i D_j K + 2 (\lambda - 1) R^{ij} R_{ij} D^2 K - 4 K^{ij} R^{kl} D_i D_j R_{kl} + 2 R K^{ij} D^2 R_{ij} \nonumber \\
&+ 4 K R^{ij} D^2 R_{ij} - 4 K^{ij} D_i R^{kl} D_j R_{kl}+ 4 K D_i R_{jk} D^i R^{jk} + 4 K^{jk} D^i R (D_i R_{jk} - D_k R_{ij}) - 2 R K^{ij} D_i D_j R \nonumber \\
&- 4 K^{ik} R_k^{\ j} D_i D_j R + 2 K R^{ij} D_i D_j R + 2 K^{ij} R_{ij} D^2 R + K R D^2 R - 2 K^{ij} D_i R D_j R + 2 K D^i R D_i R\,.
\end{align}
\end{widetext}

\end{document}